\documentclass{article}
\usepackage{bm}

\begin {document}
\noindent {\Large{\bf Significance of gyropotentials in metric theory of inertia}}

\bigskip

{\em Igor {\'E.} Bulyzhenkov}

\bigskip

						
	{Levich Institute for Time Nature Explorations},  
 {Moscow},
 {Russia}								

\bigskip
{{\bf Abstract}. The constancy of orbital velocities of peripheral stars in a spiral galaxy points to a potential regime of co-rotation together with the interstellar densities of the galactic disk.  The Einstein gyropotential rises to the evolutionary constant of a self-rotating metric profile, while the Newtonian potential falls radially. Gyropotential in vortex self-organizations creates centripetal gyroforces that prevail over Newtonian attraction of probe bodies at large distances from the axis of rotation. For practical verifications, the 1914 geodesic relations of Einstein predict the inverse distance velocities for retrograde stars. For the solar system, gyrometric corrections of Newtonian forces under transitions to  the potential circulation in material vortexes can be sought in the Kuiper Belt, the peripheral rings of Saturn and high equatorial orbits around other planets. In conclusion, the mechanical analogy with the Aharonov–Bohm effect and a possible transition to the monistic theory of metric inertia are discussed.
 }
 

\bigskip
{{\bf PACS}: 95.35.+d; 98.62.Dm; 98.62.Gq}


\section {Introduction}

Non-Keplerian orbits at the galactic periphery can be demystified by observatio\-nal study of retrograde stars in those vortex organizations where the gyropotentials of Einstein\rq{s} metric theory begin to make a significant contribution to the low-velocity limit of motion. The geodesic relations for probe bodies rely on four metric potentials \cite {Ein}, while incomplete Newtonian gravity allows only one-component potential even for rotating sources. 
To explain non-Keplerian orbits, dark matter profiles are persistently added to Newtonian physics to model the observed constancy of the azimuth velocity of stars at large distances from the axis of a rotating galaxy. 

My domestic opponents firmly support the existence of dark matter as \lq\lq a well-argued fact\rq\rq{}
 \cite{Zas}. Their instrumental infrastructure to search  for dark matter in space and underground assumes  infallibility of Newtonian mechanics for the slow motion in weak fields. It came down to a priori statements that relativistic physics is not able to get around the requirement of dark matter due to the far-fetched assumption that Einsteinian theory should allegedly repeat Newtonian one in \lq\lq the  area of joint applicability\rq\rq{} i.e. when mechanical bodies can move freely in weak accelerating fields.

The official plan of the Russian Academy of Sciences for 2006-2025 prescribed fundamental research not only \lq\lq to develop cosmological models that should take into account dark matter and seek to explain its nature\rq\rq{} analytically, but also to spend on the costly \lq\lq solution of the problem of finding the hidden baryonic matter\rq\rq{.} 
The purpose of this work is to undeline a measurable difference between Newton\rq{s} dynamics and Einstein\rq{s} geodesics in weak fields of rotating masses. And to reveal the principle Newtonian errors in agreement with Kuhn\rq{s} conceptual conclusion \cite {Kuh}: \lq\lq 
 Einstein\rq{s} theory can be accepted only with the recognition that Newton\rq{s} was wrong\rq\rq{.}
This theoretical part of the domestic discussion with dark matter proponents may be of interest to a broad readership in those  countries that were also involved in expansive searches for dark wonders.

After discussing the incompleteness of the Newtonian model in the case of vortex self-organization, an attempt will be made to reorient collective research from the dark matter highway to an informative analysis of retrograde stars and potential forms of cyclic motion in vortex distributions according to Einstein\rq{s} physics with a monistic field-matter, as in quantum mechanics. Briefly, the rotating self-organization of the galactic medium can generate three metric gyropotentials at the disk periphery in addition to the Newtonian scalar potential. With large radii of the equatorial orbits, Einstein\rq{s} gyroforces begin to supplement the centripetal forces from the inverse square law. It can be expected that possible observations of retrograde bodies in distant orbits around elliptical galaxies or even individual planets will translate Einstein\rq{s} metric formalism for the macro- and megacosm into a non-local language of the material continuum with instantaneous  correlations.

\section {Incompleteness of Newtonian gravity for fields in rotating distributions}
To make it easier to understand Kuhn and the conceptual difference between orbits in one-component and four-component potentials, we will first try to think only according to Newton, as taught in high school. One can model a spiral galaxy with a summary mass $M_\Sigma = M_b + M_d$ of a spherical bulge and a thin
disk with two-dimensional mass density $\sigma = M_d (L)/\pi L^2 = const$. Let the disk significantly go beyond the radius $l$ of the bulge containing a constant (for simplicity) mass density $\mu = 3 M_b (l) /4\pi l^3 = const$. Spherically symmetric potential
 $\varphi_b (|{\bm x}|) = - G M_b [( 3l^2 -|\bm x|^2)/2l^3 ] \Theta (l -|\bm x|) - G M_b \Theta (|\bm x| - l ) / |\bm x| $
from the homogeneous distribution of the bulge mass $M_b$ is additively superimposed in Newton\rq{s} theory on the gravitational potential
$\varphi_d ({ \rho}, { z})$ of the thin disk distribution of mass $M_d$ within the entire circle with radius $L\gg l$. 

To understand the force gradients along the plane of the disk, let us first compare the scalar potentials $\varphi_d ({ \rho} = 0, { z} \Rightarrow 0) = -2\pi G\sigma ({\sqrt {L^2 +z^2} -|z|})
\Rightarrow -2\pi G \sigma L $ in the center of the disk and on its edge,
\begin {equation}
\varphi_d (L, 0) = -G\!\!\int\! \sigma dS/r = -G\sigma\!\!\int 2\theta dr = 4\pi G\sigma L\!\!\int^0_{\pi/2}\!\theta \sin \theta d \theta = -4 G\sigma L.
\end {equation}
Here we used a geometric constraint for the integration segment from the edge of the disk ($r=2L cos\theta $, $dr = -2L sin \theta d\theta$) and integrated by parts with respect to the angular variable $\theta$.

The Newtonian potential at the center of a thin disk is smaller than at the periphery, 
since $-2\pi G \sigma L < -4 G \sigma L$. The monotonic variation of this potential along the disk radius can be approximated by the linear function $\varphi_d (0\leq \rho \leq L , z =0)
 \approx -2\pi G \sigma L + 2 G(\pi -2) \sigma \rho $.
Therefore, in addition to the radial gradients $\varphi_b (|{\bm x}|)$, the static model of the galaxy also creates centripetal forces from the gradients of the disk potential $\varphi_d$. In the simplest case of circular orbits around the axis of such a model galaxy, Newtonian physics leads to a radial balance of accelerations and specific forces in the total potential of a spherical bulge and a thin disk,
\begin {eqnarray}
\frac {v^2}{\rho} =  \partial_\rho (\varphi_b\! +\! \varphi_d) \Rightarrow  \frac {G M_b \rho} {l^3}   \Theta (l -|\bm x|)  + \frac {G M_b} {\rho^2} \Theta (|\bm x| - l ) 
+  2 G(\pi -2) \sigma.  
\end {eqnarray}   
This classical balance corresponds to Kepler orbits at moderate distances from the axis, when $\rho \ll L \sqrt { M_b / M_d }$, but on the homogeneous disk periphery it unexpectedly predicts even a radial increase in the orbital velocity $v = $ $ {\sqrt {2G( \pi -2) \sigma \rho}}$. It is clear that a uniform disk profile of interacting masses is unstable without adjusting pressure or some artificial kind of centrifugal forces. In the reality of differential auto-rotations, such a uniform profile cannot be formed either in  in statics or in dynamics. Nonetheless,  the simplicity of calculations for static profiles, such as the Mestel disk with a radially decreasing density \cite {Mes}, makes it possible to quickly assess the key points in modeling of inhomogeneous distributions.

Far from the bulge boundaries, Newtonian balance (2) could also predict the radial constancy of the orbital velocities of probe bodies in a static disk with a monotonically decreasing density $\sigma (\rho) \approx \sigma(\rho_o) \rho_o ln (e\rho/ \rho_o)/\rho$ on the far periphery of radii $\rho$, when $l \leq \rho_o \leq \rho \leq L$. But how to keep such a bulge-disk profile of a stellar medium  in statics?
Numerical simulations of spiral galaxies through inhomogeneous density distributions for scalar potentials cannot abandon the classical balance for centripetal accelerations and Newtonian forces ${v^2}/{\rho} = \partial_\rho (\varphi_b\! +\! \varphi_d) $.
Deviation from the spherical symmetry of the static profiles of galaxies without rotations would raise questions about the principles of their self-organization, albeit with dark matter.

Not only a disk, but also an ellipsoidal profile of mass-energy implies stationary motions of Newtonian sources along stable orbits.
However, there are no Lorentz-type forces in Newtonian gravity. And it is not surprising that Newton\rq{s} scalar potential failed to quantitatively match the observed luminosities in rotating galaxies with their calculated profiles. To get rid of the mystical forces of dark matter in modern physics, it is necessary to move from Newtonian gravi-mechanics with one scalar potential to the Einstein transport of probe bodies in metric space-time with 4 geodesic potentials as in Maxwell\rq{s} theory with the electromagnetic four-potential.

\section {Potential circulations of geodesic bodies in metric space-time with gyro-potentials}

The budgeting of mega-science infrastructures for the search and scientific verification of dark matter in an era of non-local quantum distributions and well-defined affine connections in the geometric organization of space-time raises many questions. It would seem that in modern mechanics it is necessary to comprehend not medieval ideas about dark forces in the surrounding cosmos, but the conceptual flaws of the dual world palliative when transferring the monistic distribution of quantum mass-energy to macroscopic scales.

Before introducing dark matter into the galactic halo of the Newtonian void, let\rq{s} recall the well-known differences between the Newton model (massless fields between massibe particles) and the mass-energy unity in quantum fields. And, let us confirm quantitatively that the nonrelativistic accelerations in the four metric  potentials of Einstein, $c^2 g_\mu \equiv c^2g_{o\mu} /{\sqrt {g_{oo}}}$, are by no means reducible to slow motion referents in the scalar potential of Newton, $\varphi_\Sigma \equiv - G\sum (m_k/r_k) = c^2{\sqrt {g_{oo}}} - c^2 \leq 0 $. Everyone knows that back in 1914, Einstein introduced the geodesic 4-acceleration
$a_\mu \equiv c^2( du_\mu/ds) -c^2{\Gamma^\lambda_{\mu\nu}} u_\lambda u^\nu \equiv c^2u^\nu \nabla_\nu u_\mu = 0$ for the free motion of the probe mass (or its scalar density $ \mu_p =const$) in arbitrary metric fields \cite {Ein}.

 In the absence of dissipation or non-metric external forces, the probe mass density (or an idealized point mass) is forced to move along the spatial coordinates, $dx_p^i \neq 0$, according to the geodesic 4-acceleration with a relativistic 4-velocity
  $cu_\mu \equiv \{ c\gamma {\sqrt {g_{oo}}} ; -\gamma [v_i - c (g_{oi} /{\sqrt {g_{oo}} }) ] \} \equiv cg_{\mu\nu} dx_p^\nu / ds $, with $u^\nu\nabla_\mu u_\nu \equiv 0$. Here, the well-known factor $c^2\gamma \equiv c^2 / { \sqrt {1 - v_iv^i/c^2}}$ is the variable kinetic potential acquired  by the probe mass density $ \mu_p $ during motion. Moreover, the physical 3-velocity $ v^i \equiv c dx_p^i /\gamma ds \equiv c dx_p^i /
 g_\nu dx^\nu$ for geodesic transfer along three spatial coordinates $x^i$ is determined in General Relativity  not by the Minkowski world coordinate  $x^o/c $, but by a specific rate of physical time
$g_\mu dx^\mu/c \equiv [{\sqrt {g_{oo}}} dx^o + (g_{oi}/{\sqrt {g_{oo}}}) dx_p^i ]/c $ for the moving density $ \mu_p $.
A motionless observer operates with the universal local time $ {\sqrt {g_{oo}}} dx^o/c$ to describe moving particles. Therefore, the locally measured velocity of any probe particle becomes  $ dx_p^i /
({\sqrt {g_{oo}}} dx^o/c)$. The physical and measured velocities coincide only under the local condition
$g_{oi}dx_p^i/{\sqrt {g_{oo}}} =0$, which emphasizes the importance of the Einstein gyropotential $c^2g_i \equiv c^2g_{oi}/{\sqrt {g_{ oo}}}$ for a correct description of the moving probe masses even in the negative potentials of  Newton  (for always positive values of the corresponding Einstein potential,
$0 < c^2 g_o \leq c^2$).
  
	In Einstein\rq{s} physics,  the free motion of the probe  scalar mass (or its local density $\mu_p c^2$) obeys the strict geodesic balances $\mu_p c^2 u^\nu \nabla_\nu u_\mu = 0$ and the metric path identities
$(-\mu_p c^2) u^\nu \nabla_\mu u_\nu \equiv 0$ for local 4-velocities $u_\mu\equiv dx_\mu/ds$ and $u^\mu\equiv dx^\mu/ds$ with the adaptive coupling $u_\mu u^\mu \equiv 1$:
\begin {eqnarray}
\cases {\mu_p c^2u^\nu (\nabla_\nu u_\mu - \nabla_\mu u_\nu) \equiv \mu_p c^2u^\nu (\partial_\nu u_\mu - \partial_\mu u_\nu) = 0 \ $or  by components:$ \cr 
 -\mu_p \gamma  v^j [c^2\partial_j (- \gamma {\sqrt {g_{oo}}}) - c\partial_o \gamma (v_j - cg_j)] \equiv - \mu_p \gamma g_o {\bm v} \cdot {\bm E} = 0, 
 \cr
\mu_p u^o \! [c^2\partial_i ( - {\gamma\sqrt {g_{oo}}})\! - \! c\partial_o \gamma\! (\! v_i\! -\! c g_i)]\! + \!\mu_p \gamma v^j
[\partial_i \gamma \!(\! v_j\! -\! c g_j)\! -\! \partial_j\gamma\! (v_i\! -\! c g_i) ]=0 
\cr
$or $ \ \mu_p u^o [c^2 {\bm \partial} (- \gamma {\sqrt {g_{oo}}}) - 
 c \partial_o ( \gamma{\bm v}\!-\!c \gamma{\bm g})  ]  + 
					   \mu_p\gamma {\bm v} \times curl [( {\bm v}\! - \!c {\bm g}) \gamma ] \cr 
		 \equiv		\mu_p (u^o g_o {\bm E} +  \gamma {\bm {\beta}} \times {\bm B} )
			\equiv		\mu_pu^o g_o [ {\bm E} +    (d{\bm x}/g_o dx^o ) \times {\bm B} ]	= 0.
}
\end{eqnarray}
 Here $ {\bm E} \equiv c^2 g^{-1}_o {\bm \partial }(-\gamma { {g_{o}}}) - cg^{-1}_o \partial_o \gamma ( {\bm v} - c{\bm g}) $,  ${\bm B(x)} \equiv
 curl(c{\bm v} \gamma) - curl (c^2{\bm g} \gamma) $, and the 3-vector $(curl \bm a)^i \equiv e^{ijk} (\partial_j a_k - \partial_k a_j)/ 2 {\sqrt {|g_{ps}|} } $ is dual to the tensor $(\nabla_j a_k - \nabla_k a_j)$. Four geodesic relations (3) for moving probe densities $ \mu_p $ generalize non-dissipative Bernoulli flows with conservation of energy along fluid streamlines ${\bm v} \cdot {\bm E} = 0 $. This geodesic conservation for the adaptive motion in metric fields with four relativistic potentials 
$c^2g_\nu \equiv c^2g_{o\nu}/g_o$ is a direct mathematical consequence of three last equations for the kinetic-metric tensions  ${\bm E}$ and ${\bm B}$.

The tension ${\bm B}$ is responsible in (3) for the relativistic imbalance between the Euler-Lagrange centrifugal force
 $ \mu_p \gamma {\bm {v}} \times curl (- {\bm v}\gamma )$  and the Einstein centripetal gyro-force $\mu_p c\gamma {\bm {v}} \times curl ({\bm g}\gamma) $  in metric organizations of vortex fields.  For the metric gyro-force, the analogy with the current components of the Lorentz force is obvious when electric charges move in an external magnetic field. Metric gyro-potentials  $c^2{\bm g}\equiv \{c^2g_1, c^2g_2, c^2g_3\}$ for probe masses were not taken into account by Euler in the classical model of an ideal fluid. And Newton\rq{s} scalar potential stayed in hydrodynamic applications not supplemented by Einstein\rq{s} gyro-potential for swirling mass-energy flows. Newtonian mechanics and Euler forces do not correspond to Einsteinian accelerations of probe bodies even in slowly rotating distribution of inertial mass-energy. Einstein\rq{s} four-potential geodesics had not been  not falsified by experiment, while the Euler/Navier-Stokes equation does not work in many applications to rotating fluids. In no case can the Newtonian dynamics with one-component potential  claim to be complete in modeling the material profiles of a vortex galaxy.

The key misinterpretation of metric spacetime after the dual 
worldview with massive particles and massless fields is that the metric field inside the galaxy disk belongs to the void with zero local inertia. Such spatial voids cannot contribute to the local creation of field gyropotentials. In 1938, Einstein and Infeld \cite{EI} proposed a non-dual reading of mechanical events in purely field terms for extremely high and low densities in the mass-energy continuum. Such  monistic worldview maintains the nonlocal organization of continuously  distributed inertia \cite{TurBul} through correlated  metric connections and  volumetric mass-energy conservations on all time intervals.

If the metric self-organization of a distributed mass-energy (called a galaxy) rotates around a fixed axis, then a motionless probe density $\mu_p$ with $dx_p^i/g_\mu dx_p^\mu = dx_p^i/g_o dx^o = 0$ and $u^o \equiv \gamma dx^o /g_\mu dx_p^\mu = 1/g_o $ will undergo according to (3) only universal metric forces
$ \mu_p (-c^2{\bm \partial} {ln g_o } + g^{-1}_oc^2\partial_o {\bm g}) $. This does not mean, as already noted, that any moving density $\mu_p \gamma_p $ will also experience only universal force actions. Moving probe bodies are subjected to very complex accelerations from the geodesic tensions (3) because of the specific potentials $\gamma_p c^2$. Thus, geodesic accelerations depend both on the universal metric potentials $c^2g_{o\mu}/{\sqrt {g_{oo}}}$ and on specific 3-velocities $v_p^i = c dx_p^i/g_\mu dx_p^\mu$ for each probe particle.

One can use the vector identity $curl \ grad\ \equiv 0$ for the metric-kinetic balance (3). This identity leads to the temporal dependence of the geodesic tension ${\bm B}$ for moving probe densities in a metric vortex,  
\begin {eqnarray}
\cases {\partial_o {\bm B} \equiv \partial_o curl [( {\bm v}\!-\!c {\bm g}) \gamma c]= 
curl [ (\gamma {\bm v} /c u^o) \times curl ( {\bm v}\gamma  - c {\bm g}\gamma  )c ]
\cr\cr
\int_{x^o_1}^{x^o_2}\! dx^o \partial_o {\bm B} \!\equiv c [curl\! ( {\bm v}\!-\!c {\bm g}) \gamma ]_{x^o_1}^{x^o_2} 
\! =\! c\!\int_{x^o_1}^{x^o_2}  curl\! \left [\frac {d{\bm x}}{d x^o }\! \times\! curl ( {\bm v}\gamma \! -\! c {\bm g}\gamma  ) \right ]\! dx^o.
}
\end {eqnarray}

Einsteinian geodesics (3) leads to mathematical consequences (4), which allow us to understand the physical meaning of the coordinate components
$c u_i = - ( { v_i} - c { g_i} )\gamma $ in the covariant 4-velocity $c u_\mu$. It can be seen from the first equation in (4) that if
$curl [ \gamma c( {\bm v} -
 c{\bm g} )]_{x^o_1}$ = 0 everywhere at the initial time $x^o_1$, then the time derivative of this value also vanishes. This means that, according to the second Lagrange theorem, $ {\bm B} (x^o > x^o_1) = 0$ at all subsequent times. The stable vanishing  $ curl [\gamma ({\bm v} - c {\bm g} )] = 0$ of the kinetic-metric tension in self-organized vortexes of matter was noticed, in particular, for the superfluid (and dissipation free) motion of helium. These steady circulations of liquid densities were called the potential motion of vortex matter.

The geometric (holonomic) organization of space-time with geodesics (3) must support potential vortexes for both weak and strong metric fields, since the balanced kinetic-metric tension ${\bm B}\equiv curl [\gamma c ( {\bm v} - c { \bm g} )] = 0$ can exist, according to (4), for a long period of the coordinate parameter $x^o$. According to Stokes\rq{s} theorem, the corresponding Kelvin circulation along any closed line can also exists during the same period of time. In general geodesic motions (3), the potential transfer of probe densities  is realized with simultaneous zeroing of both kinetic-metric tensions: ${\bm E} = {\bm B} = 0$. In this case, the steady Kelvin circulation allows a non-stationary change in the observable kinetic  fraction (or tornado),
$\mu_p \oint \gamma{\bm v}({\bf x},x^o) d{\bm l} = \mu_p c\oint \gamma {\bm g}({\bf x},x^o) d{\bm l} \neq 0$, due to compensatory changes in the Einstein gyropotential $c^2{\bm g}({\bf x},x^o)$ along the closed contour. In many isolated galaxies, the channels of dissipation of the integral mass-energy can be neglected. Then a stationary tornado of the observed matter at the periphery of the vortex galaxy can be accompanied by a finite gyropotential ${\bm g}$,  with $\oint \gamma {\bm g} d{\bm l} \neq 0$, related to the metric self-organization of galactic fields and mass-energy currents.

\section {Einstein\rq{s} gyroforces  instead of dark options}

For Keplerian  circulations of probe bodies in the central part of the galaxy and in its near periphery, a non-potential kind  of non-relativistic motion is realized, when one can set in the geodesics (3) that  ${\bm v}({\bm x}, x^o) \approx  c d {\bm x}/dx^o, $ $v^2 \ll c^2 $,   $ {\bm B} \approx curl ( c{\bm v}  - c^2 {\bm g}) \neq 0$,  
$\partial_i \gamma \approx \partial_i v^2/2c^2$, ${\bm g}^2 \ll 1$, $c^2 (1-{\sqrt {g_{oo}}}) = -(\varphi_b + \varphi_d)\equiv -\varphi_{_\Sigma}  \ll c^2 $, ${\bm E} \approx - {\bm \partial} ( \varphi_{_\Sigma} + v^2/2 ) - c\partial_o (\! {\bm v} - c {\bm g}) \neq 0$. 
Let us consider the simplest case of a common axis of symmetry $z$ for metric fields and circular orbits of probe masses, when
${\bm g}(\rho, z) = {\hat \alpha} g (\rho,z)$, and ${\bm v}(\rho, z) = {\hat \alpha}v(\rho,z) $. Then, specifying the unknown metric potentials in general form as $c^2g_o(\rho, z) = c^2 + \varphi_{_\Sigma}$ and $c^2g_i(\rho,z) = \{0, c^2g(\rho,z), 0\}$, one can start searching for orbital velocities for the nonrelativistic limit in the Einstein geodesic (3).

Kepler\rq{s} equatorial orbits in the $z=0$ plane are supported by two centripetal forces $- m_p{\hat {\bm \rho}}{\partial}_\rho v^2/2 = - m_p \omega^2 {\bm \rho } $ and $- {\hat {\bm \rho }} m_p {\partial}_\rho \varphi_{_\Sigma}$ (from the kinetiv-metric tension ${\hat {\bm \rho}}E_\rho \neq 0$) counteracting the Euler-Lagrange centrifugal force $ m_p {\bm v} \times curl {\bm v} 
= + 2 m_p\omega^2  {\bm \rho} $. This dynamic confrontation of radial forces from non-potential geodesic tensions leads to the vector sum $- m_p\omega^2 {\bm \rho}$ and the resulting acceleration $- \omega^2 {\bm \rho}$ towards the center of the Kepler orbit.
Einstein\rq{s} centripetal gyroforce $  m_p  {\bm v} \times curl(- c{\bm g})$ cannot compete in (3) with Newton\rq{s} gravitational force $ {\hat {\bm \rho }}m_p(-{\partial}_\rho \varphi_{_\Sigma})$ at small and medium radial distances.

In the rotating Solar system, for example,  mass-energy densities are grouped almost spherically near the origin, and the radial growth of the relatively weak gyropotential $c^2g (\rho,z)\ll c^2$ cannot strongly perturb the Keplerian rotations of the first nine planets obeying the finite  tensions ${\bm E} \neq 0$ and ${\bm B} \neq 0$ in the geodesics (3). The trans-Neptunian area and the Kuiper Belt at 30–55 AU from the Sun is practically not studied because of the low luminosity. In disk galaxies, well-observed peripheral stars co-rotate together with the interstellar medium at kiloparsec distances, where the differential rotation of the mass-energy profile brings the evolutionary gyropotential to the saturation level. This equilibrium  gyropotential at the far periphery of a spinning galaxy can correct and even completely compensate for the Euler-Lagrange centrifugal force,
$m_p {\bm v}\times curl ({\bm v} - c {\bm g}) \rightarrow 0$:
   \begin {eqnarray}
\cases 
 {	 c \partial_o ( {\bm v} - c {\bm g} ) +     {\bm \partial}
 [(\frac{v^2}{2}) +  c^2({\sqrt {g_{oo}}} - 1) ]   = 
					   {\bm v}\times curl ( {\bm v} - c {\bm g} ),
																\cr \cr 
	   \partial_\rho
 [\frac {v^2(\rho)}{2} +  \varphi_{_\Sigma}]   = 
					    \frac {v(\rho)}{\rho} \partial_\rho[\rho v(\rho) - c\rho g(\rho)],
							\cr \cr \
							v^2(\rho) = \rho \partial_\rho \varphi_{_\Sigma} + v(\rho) c\partial_\rho [\rho g(\rho)], 
							\cr\cr
							v_{1,2} = \frac{1}{2}\partial_\rho (c\rho g) \pm 
							\frac{1}{2}{\sqrt { [\partial_\rho (c\rho g)]^2   + 4\rho \partial_\rho \varphi_{_\Sigma}  } } =\cr\cr 
							\cases { 
							{\sqrt { \rho \partial_\rho \varphi_{_\Sigma}  } } + \frac {c}{2}\partial_\rho (\rho g) \ 
							$and$ \  
							\frac {c}{2} \partial_\rho (\rho g)- {\sqrt { \rho \partial_\rho \varphi_{_\Sigma}  } },  \ $if$ \   [c\partial_\rho (\rho g)]^2 \ll {{ \rho \partial_\rho \varphi_{_\Sigma} } }      \cr\cr 
				c\partial_\rho (\rho g)	+ 		  \frac {\rho \partial_\rho \varphi_{_\Sigma} } {c\partial_\rho (\rho g) }   
				\ $and$ \ - 	\frac {   \rho \partial_\rho \varphi_{_\Sigma} }{ c\partial_\rho (\rho g)    },  
 				 \ $if$ \ {{ \rho \partial_\rho \varphi_{_\Sigma}  } } \ll  [c\partial_\rho (\rho g)]^2     
							}.
}\end{eqnarray}

Strong radiative losses or intensive gas exchange with the intergalactic medium can lead to non-stationary changes in the Newtonian potential for peripheral orbits  due to temporal changes in the inter-orbit mass integral. In such a non-stationary galaxy, the temporal derivatives of $\partial_o \{...\} $ can give additional centripetal accelerations to perturb Keplerian orbits due to the intervention of dissipation derivatives into Newton\rq{s} gravimechanics. Now, we neglected evolutionary dynamics in (5) for the orbital velocity solutions $v_{1,2}$,  by considering the stationary case of motion with $\partial_o ( {\bm v} - c {\bm g} )=0$.

In contrast to the radially decreasing gyropotential in the Kerr metric $g_{Kerr}$ $\propto const /\rho^2 $ for the massive singularity with spin \cite {Ker}, we admit the massive medium in the rotating disk periphery  where $\rho \partial_\rho \varphi_{_\Sigma}/c^2$ can be compared with $ [\partial_\rho
(\rho g)]^2 \approx const \ll 1$. Then Keplerian rotations can be perturbed in (5) due to local damping of the Euler-Lagrange centrifugal force by the Einstein centripetal gyro-force. The post-Keplerian regime of orbital circulations in the Einstein geodesic starts at $ (c \partial_\rho [\rho g(\rho)])^2 \geq 4\rho \partial_\rho \varphi_{_\Sigma}$, which allows us to estimate from observations  the peripheral gyropotential for the most of disk galaxies as $g \approx 10^{-3} - 10^{-2}$ .

In slowly circulating distributions of galactic mass-energy, the gyropotential increases monotonically with distance from the rotation axis, but should reach its limiting saturation at non-relativistic values, $g (\rho \geq \rho_*, z=0) \Rightarrow g_* = const \ll 1$. After this saturation, the orbital speed of  disk densities cannot grow with radius and remains small compared to the speed of light.
In principle, the two-variable  function $g(\rho,z)$ must be consistent with the metric solutions $ {\sqrt {g_{oo}(\rho,z)}}$ for the field - matter organization in the  inertial profile of a rotating disk. This metric self-organization has not yet been satisfactorily described a rotating medium with constant mass-energy and spin integrals due to non-locally organized affine connections. However, non-relativistic geodesic (5) and astronomical observations of retrograde stars in vortex galaxies together can shed new light on the spatial gradients of the metric 4-potential $c^2 g_{\mu}\equiv c^2 g_{o\mu}/ {\sqrt {g_{oo}}}$ in rotating self-assembly of inertial matter.

\section {Potential motion on high orbits around elliptical galaxies and planets}

The Newtonian potential beyond  an elliptical galaxy can be approximated in the equatorial plane by the textbook function  $\varphi_{_\Sigma}(\rho, z=0) = - GM/\rho$. Can one expect the geodesic curve to be constant beyond the visible frames of compactly rotating  systems, as was claimed from (5) for the massive disk medium of a spiral galaxy? 

Geodesic relations (3) at very large radial distances can always have a stationary solution ${\bm B} = curl ({\bm v} - c {\bm g})c \Rightarrow 0$ and
${\bm E} = -c^2 \gamma{\bm  \partial}ln (\gamma g_o) \Rightarrow 0$ for potential circulations around a slowly rotating galaxy, since it always generates a finite, albeit very small, gyropotential ${ g(\rho \gg \rho_*) \approx const \equiv g_* }$.  We try the orbital velocity for such potential circulations in the $z=0$ plane as follows:  $ {\bm v}(\rho, z=0)/c = {\hat {\bm \alpha }}{g}(\rho,z=0) + a[{\hat z}\times {\hat \rho }]/\rho $ or $v (\rho) = cg_* + c(a/\rho)$ for $\rho \geq \rho_*$. The so-far undefined scale $ a=const$ will be specified by nonrelativistic geodesic relations:
\begin {eqnarray}
\cases { 
 curl [{\bm v}(\rho) -c {\bm g}(\rho)]  = 0 , \ 
v (\rho) = cg_{*} + c\frac {a}{\rho}  \ll c, \cr
{\bm \partial} [\varphi_{_\Sigma}(\rho) + \frac {v^2(\rho)}{2}] = 0,\ 
2\varphi_{_\Sigma}(\rho) + c^2[g_{*} + \frac {a}{\rho}]^2 =  v^2(\infty) = c^2g^2_{*},
\cr  g_{*} =  \sqrt {g_*^2 + \frac {(- 2\varphi_{_\Sigma})}{c^2}} - \frac {a}{\rho} \approx g_{*} + 
\frac{(-\varphi_{_\Sigma})}{c^2 g_*} - \frac {a}{\rho}, \cr
a = \frac {(-\varphi_{_\Sigma})\rho}{g_{*}c^2} \Rightarrow \frac {GM}{g_{*}c^2} = const,
\ \ 
\frac {GM}{c^2\rho} \ll   g^2_{*} \ll  1   
}\end{eqnarray}
 
This potential solution $v(\rho) = c g_* + ({GM}/{g_{*}c \rho}) \approx cg_* = const$  acquires the constancy of radial curves at $\rho \gg {GM}/{g^2_{*}c^2} \equiv \rho_*$. The high orbit velocity $+ {\hat {\bm \alpha }} cg_*$ in the $z=0$ plane corresponds to the angular co-rotation of the visible galaxy and probe bodies. A retrograde body in such high orbits must have the backward velocity ${\bm v}_b(\rho)\equiv {\hat {\bm \alpha }}\rho \omega (\rho)  = - {\hat {\bm \alpha }}GM/c g_* \rho$ from (5) with $\partial_\rho (g\rho) \Rightarrow g_*$ and $\rho \partial_\rho \varphi_{_\Sigma} \Rightarrow GM/\rho$ for the textbook potential of an elliptical galaxy. The backward motion is not  potential, since ${\bm B} = -{\hat {\bm z}} g_*/\rho  \neq 0$ and ${\bm E} = -{\hat {\bm \rho}} GM/\rho^2 \neq 0$ in the geodesic relations ${\bm E} +  {\bm v}_b \times {\bm B} = 0$ for the high equatorial orbits  at $\rho \gg \rho_*$.  Needless to say, that these  geodesic relations also maintain  the Kepler law $ v^2(\rho) \equiv \rho^2 \omega^2 (\rho)  = GM / \rho $ for the low orbits at $\rho \ll \rho_*$.

	Potential circulation in the $z=0$ plane can be realized not only for perepherial orbits around elliptical galaxies, but also for high equatorial orbits around rotating planets. Here the equatorial gyropotentials also acquire finite, albeit small, values $c^2 g(\rho \rightarrow \infty ) = const \neq 0$. The metric fields of the spinning mass-energy are inhomogeneous along the $z$ axis, which significantly complicates their relativistic calculation. Nevertheless, upon transition to the potential regime of stationary circulations in high equatorial orbits, one should expect $ {\partial}_z [\varphi_(\rho,z) + {v^2(\rho,z)}/{ 2 }] = 0 $ from the potentiality condition ${\bm E}=0$. Such a potential motion requires gradient forces along the $z$ axis, which push the circulating bodies toward the $z=0$ plane. These gradient forces arise even for the middle orbits, where $\rho \approx\rho_*$, by making, for instance, the peripheral rings of Saturn\rq{s} disk extremely thin.

 Destroying the nonlocal organization of gyrometic fields for potential circulations in the disk of Saturn by local mechanical influences is apparently  as difficult as destroying the nonlocal self-organization of a tornado in the atmosphere  or Kelvin circulation along closed macroscopic contours in an ideal fluid. One could expect non-dissipative reflection of radio waves by inert bodies on the potential orbits, anomalously slow relaxation of disk dislocations after collisions with external asteroids, and suppression of paired interactions between potentially circulating bodies on neighboring rings of  Saturn\rq{s} disk. For example, the co-orbital moons Janus and Epimetheus have roughly equal masses of about $10^{18}kg$, but their orbits (only a few kilometers apart along the semi-major axis) do not result in the mutual collision expected from Newtonian gravity. Potential motion in Einstein\rq{s} vortex fields allows only slight fluctuations of orbits, which are observed every four years \cite{Spi}.

The geodesic relations ${\bm E}+ ({\bm d{\bm x} }/g_odx^o)\times {\bm B} = 0$ are applicable to any orbits of the Solar System, but the potential case $ {\bm E}= {\bm B}=0$ should only be expected at the far periphery of metric vortexes, including the Kuiper Belt. It can also be recommended from (5) to search for the radial constancy of the azimuthal velocities of co-rotating satellites and the inverse radial curves for the backward velocities of retrograde satellites in different  high orbits around Jupiter, Uranus, Neptune and other spinning concentrations of inertia.  Such a directed search will help to abandon the dark matter hypothesis and to justify Einstein\rq{s} geodesics for the \lq anomalous\rq{} rotation in cosmology.

\section {Conclusion}

The main conclusion from the strict mathematical revisit of the relativistic geodesics  is that the non-relativistic limit for metric forces and paired Newtonian interactions give different predictions for observations. Fields with a 4-component potential and a one-component potential should lead to measurable differences in the practice of vortex motions. Moreover, Einstein\rq{s} theory of General Relativity quantitatively and qualitatively differs from Newton\rq{s} physics even at zero speeds of spatial translations because of the internal degrees of kinetic freedom (with variable rest-energy deposits under the spatial transport).  After all, Einstein\rq{s} heated stone should weight more, while Newton\rq{s} has the same weight.

Kuhn has been right indeed in saying that \lq\lq  Einstein\rq{s} theory can be accepted only with the recognition that Newton\rq{s} was wrong\rq\rq{.} In Einstein\rq{s} metric description of the united energy-momentum there is no independent concept of negative (gravitational) energy, but there are coordinate shifts of the unexposed kinetic  fraction $mc^2$ (which can be extracted, for instance, by nuclear explosions). The total energy content of any probe mass $m_p$ is not a dual sum of kinetic, $m_pc^2\gamma > 0 $, and gravitational, $m_p\varphi_\Sigma  \leq 0$, contributions of different nature, but the monistic whole of unexposed and exposed  kinetic contributions to the always positive product, $\mu_pc^2 {\sqrt {{g_{oo}}/ (1-\beta^2) }} > 0$. There is only one (kinetic) four-vector composition for the  energy-momentum of mechanical bodies in General Relativity. And there are no  independent four-vector compositions in co-variant transformations to support the dual concept of kinetic and potential entities.   

Einstein\rq{s} metric counterpart of Newton\rq{s} scalar potential
 is always and everywhere positive
$c^2{\sqrt {{g_{oo}(x)} }} \equiv c^2 + \varphi_\Sigma (x)> 0 $, even for strong external fields. This geometric potential   changes only positive values of the monistic (kinetic) energy along the body trajectory.  At the same time, Einstein\rq{s} metric tension are everywhere local in confirmation of their kinetic  (pushing) nature in  the geometric organization  of holonomic space-time.

Not only Plato, Aristotle, Descartes, Hegel, Mi, all Russian and Eastern cosmists proposed to interpret mechanical phenomena in monistic terms of material fields, but also Einstein and Infeld called for a non-dual theory of the material field \cite{EI}. Everyone understands that the motion of mass-energy densities determines metric fields. And that the local responses of metric fields or their tensor stresses can also determine the motion of densities themselves. Further development of such relativistic ideas about instantaneous feedback for the adaptive correlation of mechanical motion and the corresponding metric fields will fail in the dual concept of localized masses and distant fields in void. An advanced metric theory of inertial densities does not have to follow the Newtonian declarations, but must self-consistently calculate an adaptive geometry of space-time for a rotating mass-energy distribution, i.e. with correlated affine connections and metric components at all values of the world parameter $x^o$.
Indeed, the measured conservation of the system integrals at different times requires strict nonlocal correlations in the corresponding distributions of mass-energy, momentum, and angular momentum. In problems of metric self-organization of a rotating galaxy, what is of interest is not the Kerr field for  emptiness  (presumably far somewhere from a rotating mass), but a self-consistent geometry for an adaptive mass-energy medium without  voids for its inertia.  

If the peripheral coexistence of gyrofields and intense mass currents in the visible galactic disk could somehow save the dual model (as in the empty-space electricity of point charges), then the constancy of orbital velocities around an elliptical galaxy or planet suggests to assume invisible mass-energy currents even on the very high orbits.
Here centripetal giro-forces should be supported by local \lq displacement currents\rq{}, but not   
Newtonian void, which cannot circulate at all. The existence of potential motion around spinning galaxies and planets, as well as the corresponding retrograde anomalies in mechanical systems,  duplicates the Aharonov-Bohm phenomenon \cite{AB} of material circulations along the electromagnetic 3-potential instead of presumed Newtonian void beyond the solenoid.

In general, the observed constancy of cosmological curves and the predicted inverse curves of retrograde circulations can substantiate the local inertia of material space in monistic physics on mega and macro scales in the same nonlocal way as in the quantum microcosm. The metric theory of continuous inertia in such matter-space with the adaptive local time for macroscopic distributions of non-local mass-energy can also be checked or falsified in the laboratory due to the appropriate modification \cite{TurBul} of the Euler/Navier-Stokes dynamics. The monistic theory of metric inertia (but not gravity) justifies the non-local nature of turbulence and offers high-order temporal derivatives for non-relativistic fluids  in simple laboratory probes. The non-Newtonian motion  of inertial space along its gyro-potentils implies different energy consumption for secondary restarts of gyroscopes compared to the first launch. Less energy may be required to restart in the same direction of rotation,  and more - in the opposite direction.

The dual theory of many bodies and their gravitational field encountered unresolved difficulties in calculating the self-consistent profile of a galaxy with differentially rotating substance. Overcoming these difficulties is more promising in the monistic theory of an adaptive metric field, which, due to its non-local affine connections-correlations, regulates the equilibrium distribution of rotating mass-energy fluxes according to the hydrodynamic scenario for non-homogeneous continuum of purely kinetic energy (with thermal  and ethereal  fractions of inertia). Here, retrograde stars can provide an important tool for further understanding of monistic physics for material space with nonlocal metric vortexes in cosmology, the solar system, and the earth laboratory. By closing, the methodical transition from the dual model of separated particles and fields to material fields and waves with quantization in a continuous distribution of non-local mass-energy deserves more attention than the costly search for dark matter in the emptiness of Newtonian space.


\end {document}